# Detection of chiral domains and linear electro-optical response in planar-aligned cells of an achiral rod-like ferroelectric nematic liquid crystal

Neelam Yadav[1], Yuri P. Panarin[2], Jagdish K. Vij[1*], Wanhe Jiang[3], Georg H. Mehl[3*]

[1]Department of Electronic and Electrical Engineering, Trinity College Dublin , The University of Dublin, Dublin 2, Ireland
[2]Department of Electrical and Electronic Engineering, TU Dublin, Dublin 7, Ireland
[3]Department of Chemistry, University of Hull, Hull HU6 7RX, UK

**Abstract:** Chiral domains of opposite chirality are observed surprisingly in a nematic phase of an achiral rod-like liquid crystalline mesogen that shows ferroelectric nematic and antiferroelectric smectic phases. The observations of chiral domains of opposite chirality are enabled by application of a weak electric field across a planar-aligned cell. However once domains of opposite chirality are created, these stay permanently even if the field is removed. The observed phenomenon is due to the symmetry breaking of achiral mesogens of rod-shaped molecules. These domains are similarly observed in the $N_x$ phase. However for temperatures close or in the $N_F$, the size of domains is reduced significantly below the wavelength of visible light. There is either a single domain with dipole moments parallel to the surface or domains antiparallel in the $N_F$ phase, not possible to determine in this experiment.

*jvij@tcd.ie; g.h.mehl@hull.ac.uk

**Introduction:**

Following the predictions made by Born [1] over a century ago the ferroelectric nematic phase [2, 3] has recently been discovered in rod-like mesogens where large dipole moment groups, such as fluoro or nitro groups are attached strategically to the aromatic parts of the molecule. The observation and confirmation of the ferroelectric nematic phase ($N_F$) has opened new directions and applications for liquid crystals. We need to understand the novel nematic-nematic phase transitions being currently observed with a view to not only to advancing the basic science of this discovery but also in harnessing potential for the applications of this phase. Using this technology, possible applications could include development of supercapacitors with variable and stable capacitance for use at ac voltages in energy storage devices, new types of sensors and fabrication of new devices for use in photonics. A range of applications of the $N_F$ phase may be far extensive than for the ferroelectric smectic due to a fluid nature of the nematic phase. The characteristic properties of the ferroelectric nematic need to be determined and relationship drawn in terms of the structure and properties. Furthermore, the mechanisms responsible for the formation of the multiple nematic phases are yet not fully understood either. The first step in determining and understanding the structure-property relationships is to examine the varied facets and characteristic properties of the ferroelectric nematic using a range of complementary physical techniques. Firstly, we focus our studies using a generic material, called DIO here. Once techniques used to exploring varied facets of the phase are established and the properties determined, these can then be tuned with the molecular structure. A lack of understanding of the ferroelectric nematic is proving to be a road-block for technological developments and applications of the ferroelectric nematics at present.

**Results and Discussion:**

In this paper the nematic (N), $N_x$ and $N_F$ phases of a rigid achiral rod-like mesogen are being investigated. The structure of the compound DIO with its transition temperatures are shown in Figure 1. A number of dipoles with higher dipole moments are attached to the various parts of the molecule. The liquid crystalline compound is studied in planar and homeotropic aligned liquid crystalline cells. The phase sequence observed is Iso - N ($M_1$) – $N_x$ ($M_2$ or $SmZ_A$) - $N_F$ ($M_P$) [2], in which the phase Nx is unknown or $SmZ_A$ phase as proposed by Clarks group recently [4] and $N_F$ is ferroelectric nematic. A range of techniques including polarizing optical microscopy (POM), electro-optics and dielectric spectroscopy are used to investigating the characteristics of the fascinating material in its various phases.

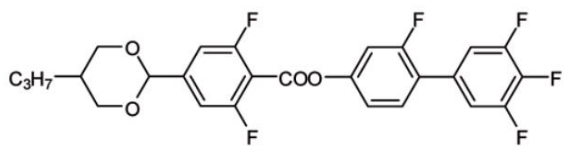

On cooling: Cr 34 MP (N$_F$) 68.8  M2 (Nx or (SmZA)) 74.5  M1 (N) 190  Iso  (˚C)

Figure 1. Molecular structure and phase sequence displayed by DIO [4, 5].

The compound was first synthesized by Nishikawa et al. [5] and has been resynthesized. However, most of the studies so far have focussed on the M$_2$ (or SmZ$_A$) and M$_P$ (or N$_F$) phases. We investigated this compound and specifically focussed to the higher temperature N phase as well. Figure 2 depicts textures of DIO in a 15 µm thick planar-homogeneously aligned cell at different temperatures. Homogeneous and uniform textures change colour as temperature is lowered from the Iso phase. A change in the colour is indicative of the change in birefringence. As temperature of the cell is lowered, fluidity of the image in terms of fluctuations is suppressed, and textures start changing from uniform to a non-uniform one with polydomains emerging in the N$_x$ and N$_F$ phases as well.

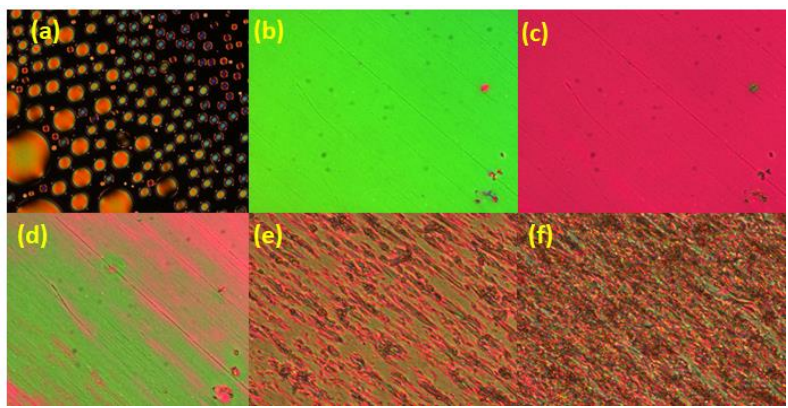

Figure 2. Planar-aligned 15 µm thick cell of DIO. Textures (a) at the transition from I to N phase (b, c) at 160 ˚C and 130 ˚C in the N phase (d) at 82 ˚C in the N$_x$ (or SmZ$_A$ phase) (e) at 66˚C close to the N-N$_F$ phase (f) at 60 ˚C in the N$_F$ phase.

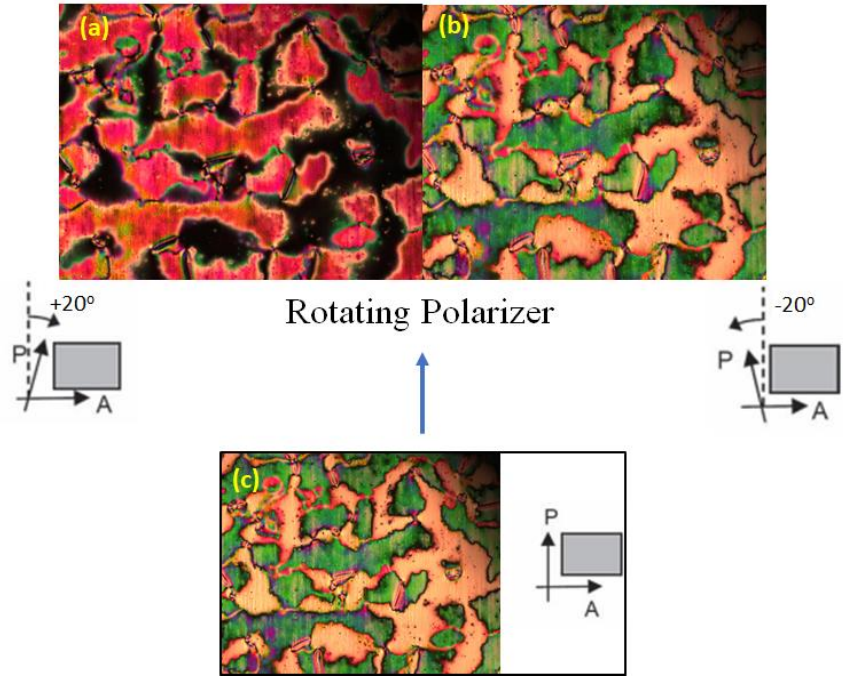

Figure 3: Textures observed in the N phase (5 V across 15 µm thick cell) at T=150 ˚C of planar-homogeneously aligned cell (c) the image recorded under crossed polarizers whereas images (a, b) are observed between slightly uncrossed polarizers showing dark and bright domains. These observations are indicative of the presence of domains of opposite chirality [6].

On the application of the electric field (0.33 V/µm) across a planar-aligned cell at temperatures within the N phase and close to the $N_x$ phase, the polarising optical microscopic (POM) images show chiral polar domains (Figure 3). The birefringent texture is composed of distinct domains observed under cross polarizers (Figure 3c). Uncrossing the polarizers by a small angle (20˚ shown in the figure) in one direction leads to the emergence of the bright and the dark domains. While, uncrossing the polarizer in the opposite direction, brightness of these domains (Figure 3a, b) is reversed. These observations prove that the distinct domains observed here under weak field are chiral and are of opposite chirality. It may be noted that the domains though enabled by the field remain permanently even if the field across the cell is removed. As the temperature is lowered and for temperatures in the $N_F$ phase, smaller domains of varied colours tend to appear. It becomes difficult to distinguish chirality or polarity of domains at the lower temperatures. Large values of the polarization density are found in the $N_x$ phase, this increases and finally saturates at lower temperatures within the $N_F$ phase. The polar nature of the N and $N_X$ phases however are confirmed by the dielectric measurements; the results are presented in a separate paper [7].

The results obtained from the POM, electro-optical measurements show a strong dependence on the thickness of the cell and temperature. Unlike the ordinary nematic, the N phase in this compound exhibits unique and fascinating properties not known before. These include the linear EO response (output/input) for the first harmonic (see Figure 4). The first harmonic signal is not observed in a conventional nematic phase, where only the output dominates for the second harmonic.

In this work, output signals for both the first and second harmonics are observed in the N phase in this material. The linear electro-optic response at a temperature of 145 ˚C is frequency dependent and decreases with increase in the frequency of the field. This shows that molecular dipoles in the N phase contribute to it and these start lagging behind alterations of the electric field as the frequency of the external signal is increased. The electro-optical observations at a temperature of 110 ˚C are frequency selective and related possibly to the pre-transition effect in the $N_x$ phase.

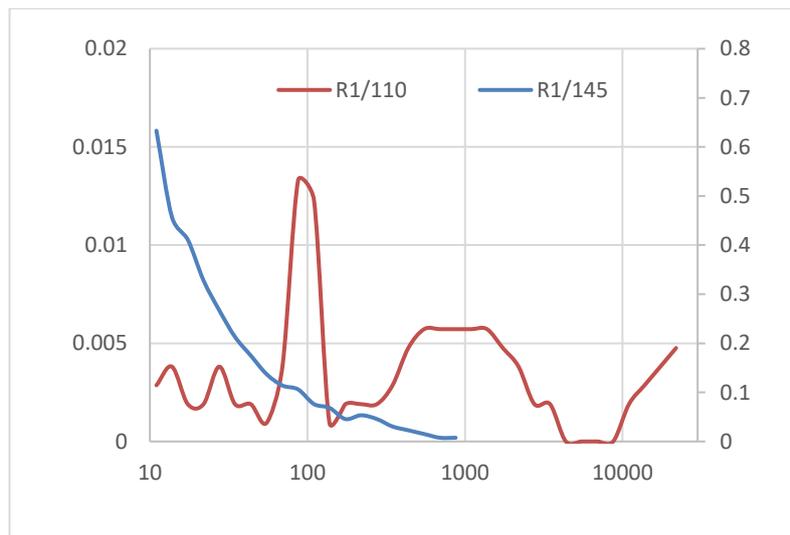

Figure 4: Frequency dependence of the linear EO response at temperatures of 110 ˚C and 145 ˚C both in the N phase. Response at a temperature of 110 ˚C closer to the $N_x$ phase is frequency selective.

At lower temperatures in the $N_x$ phase, these domains are observed too. However, in the $N_F$ phase, the size of the domains is significantly reduced, thus it was not possible to characterize the domains as in the $N_x$ phase. Observations and the physical nature of the phenomena need to be examined in terms of (i) large molecular polarizability of the medium, (ii) polarity/chirality of macroscopic domains, and (iii) the dynamic distribution and optically activity of domains.